# Photo-induced carrier dynamics in InSb probed with broadband THz spectroscopy based on BNA crystals

Elodie Iglesis[1], Alexandr Alekhin[1], Maximilien Cazayous[1], Alain Sacuto[1], Yann Gallais[1] and Sarah Houver[1]

1. Université Paris Cité, Matériaux et Phénomènes Quantiques UMR CNRS 7162, Bâtiment Condorcet, 75205 Paris Cedex 13, France

We report an optical pump - terahertz (THz) probe study of the photoinduced transient carrier dynamics in the low bandgap semiconductor Indium Antimonide (InSb). Using an organic N-benzyl-2-methyl-nitroaniline (BNA) crystal as a broadband THz source, we access the full spectral response over more than 5 THz, for varying pump-probe delay following the optical excitation. Using the Drude-Lorentz model accounting for differences between the excited length in material and the penetration depth of THz beam in pumped InSb, we extract the absolute carrier density as a function of the pump-probe delay, and provide insights on the diffusion length at given carrier densities, for different pump fluences. The mismatch between the THz penetration depth and the actual excited sample depth after carrier diffusion is discussed, since their evolutions with time and pump fluence are not intuitive as both quantities depend on carrier density.

1. Introduction

THz time domain spectroscopy (THz-TDS) has become a widely applied technique to study various materials, ranging from polymers[1] to quantum materials[2]. Since it measures the THz electric field (and not its intensity), THz-TDS provides both amplitude and phase information, and hence gives access to both real and imaginary parts of the complex optical functions, without using Kramers-Kronig relations. THz-TDS relies on a measurement of the temporal evolution of THz electric field, $E_{THz}(t)$, where one optical pulse is used to generate the THz emission and a second optical pulse (typically < 100 fs) called the gate pulse, samples the THz field as a function of time t. The THz-TDS can be easily integrated in a pump-probe scheme, such as in optical pump - THz probe (OPTP) spectroscopy, to study transient photoconductive properties of semiconductors and their carrier dynamics[3,4] and more recently photo-induced dynamics in emerging materials[5]. There, a rather intense optical pulse is used to excite the sample, and the THz pulse probes the state of the system at different pump-probe delays $\tau$.

Two approaches have been carried out for above-gap photo-excitation of semiconductors. The fastest and most common approach imposes a fixed gate delay, usually at the maximum of the THz field amplitude, and records the THz amplitude at this gate time, as a function of pump-probe delay. The output is then the relative transmitted/reflected THz field $\Delta E_{THz}(\tau)$, which is assumed to be proportional to the photo-induced carrier density and can usually be fitted by exponential models to extract relaxation and recombination time constants. This approach neglects photo-induced changes of the temporal shape of the THz waveform and hence any frequency dependence is missed. Early arguments that OPTP analysis was non-trivial and could lead to artefacts can be found in ref [6].

The second approach relies on exploiting the THz spectral response for different pump-probe delays, which requires to measure the full THz temporal trace, scanning the gate time, for each pump-probe delay. Analyzing the spectrum, one avoids the artefacts from temporal shifts or modification of the THz waveform due to a change of spectral content. Moreover, the spectral response can contain quantitative signatures of the carrier density, such as the plasma frequency, that can be modelled to extract their evolution as a function of the pump-probe delay. The spectral approach becomes even more essential with the generalization of broadband THz sources (> 3 THz bandwidth) as THz probe, such as air-plasma[1], spintronic emitter[7], THz shockwave[8], MNA crystals[9] and BNA crystals (the present work). Finally, with the advent of fast delay lines and data acquisition systems, the drawback of this approach due to long measurement time, tends to disappear.

In this work, we performed OPTP measurements on a sample of semiconductor Indium Antimonide (InSb) using a N-benzyl-2-methyl-4-nitroaniline (BNA) crystal[10] as a broadband THz source with more than 5 THz bandwidth. If THz generation from BNA crystals has shown high field[10] or high average power[11], to our knowledge, it is its first implementation as THz probe for an actual material spectroscopic study. We investigated the transient carrier dynamics induced by a 1030nm-pump, by measuring the time-resolved THz reflected spectra on a bulk sample of InSb. It is a narrow-bandgap semiconductor with the room temperature bandgap energy $E_g = 0.17$ eV. It has been well studied by THz spectroscopy in equilibrium[12,13] and out of equilibrium following the pump excitations in the optical[8,14,15] and in the THz[16,17] spectral ranges. Due to the low effective mass and the high mobility of the charge carriers, their dynamics is quite complex and includes pump-induced impact ionization and intervalley scattering[18], and carrier recombination through Auger and Shockley-Read-Hall processes[19], over rather long recombination time, lasting from nanoseconds (ns) to microseconds (µs)[20]. In previous OPTP experiments performed on InSb, two approaches have been employed: with a fixed gate delay measuring the transmission through thin films[14,21] or analyzing broadband spectral response from InSb, but at fixed pump-probe delay[8]. To our knowledge the pump-probe delay dependence of the spectral response of InSb has not been addressed on such a broadband probe range.

Here, relying on spectral analysis, we extract the carrier density for several pump fluences, as a function of the pump-probe delay from the Drude-Lorentz model, accounting qualitatively for carrier dependent penetration-depth mismatches between the excited part of the material and the THz probe. Our model also provides insights on the carrier diffusion length, which becomes comparable or even smaller than the penetration depth, depending on the carrier density.

2. Experimental setup and sample

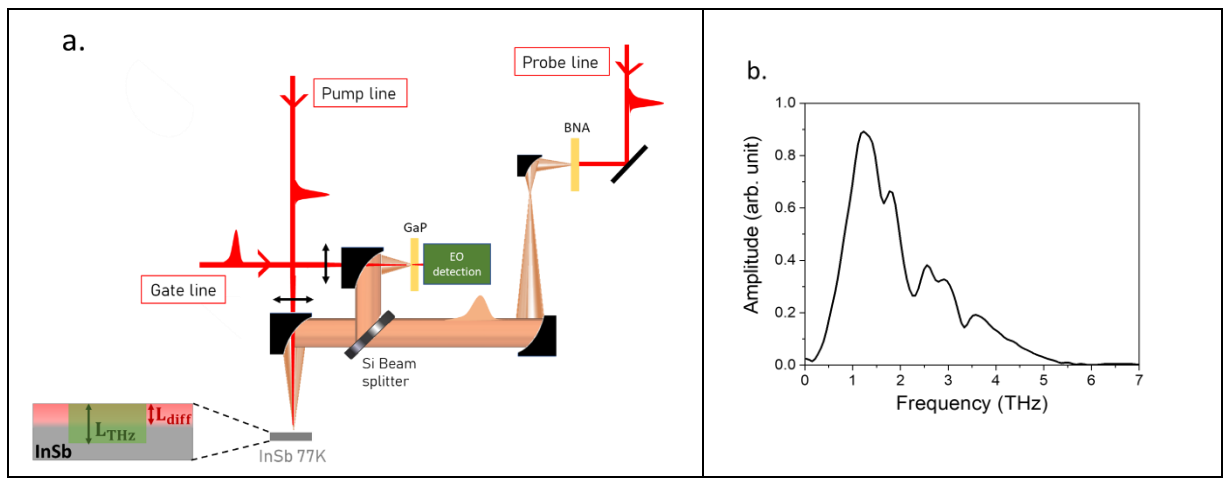

Figure 1: a. OPTP experimental setup. The laser is split in three lines: (i) the pump pulse excites the InSb sample, (ii) the probe pulse excites the BNA crystal to generate the THz pulse, which is then focused on InSb and (iii) the gate pulse is focused on a 1mm-thick GaP crystal for the electro-optic detection. A zoom on the sample schematizes the photo-excited part of the sample after carrier diffusion $L_{diff}$ and the THz penetration depth $L_{THz}$, which could be different from $L_{diff}$. b. THz spectrum generated from BNA crystal, measured in reflection as the reference.

We implemented a time-resolved THz spectroscopy set-up with an optical pump. A scheme of the set-up is shown in figure 1a. We used a 1030nm-Yb:YAG laser, generating 480 fs-pulses at 1 kHz repetition rate. Optical pulses were nominally compressed down to 60 fs, using a multi-pass cell compressor. The optical beam is divided into three paths: the optical pump, the THz probe and the gate.

The optical pump beam is telescoped and collimated to a spot size of 0.93 mm $1/e^2$-diameter on the InSb sample placed in the THz focal plane. The pump spot was set to be twice larger than the probe spot. The power was adjusted using absorptive filters to change the fluence between 8 µJ/cm², 17µJ/cm² and 31 µJ/cm² for the pump-probe measurements, and 310 µJ/cm² for the reference measurement. The optical pump excites carriers in the region close to the sample surface which then diffuse on a depth determined by the carrier diffusion length $L_{diff}$.

The THz pulses are generated via optical rectification in a ~400µm-thick BNA crystal from TeraHertz Innovations[22] which has been fused to a 0.5mm-thick sapphire plate. The optical beam is collimated and has an energy of 35 µJ/pulse, leading to a fluence of ~ 3 mJ/cm² on the BNA crystal. The spectrum extends up to more than 5 THz, as shown in figure 1b. The THz beam is directed and focused with off-axis parabolic mirrors on the InSb sample, with a spot size of 0.54 mm $1/e^2$ diameter. The THz electric field penetrates in the sample over a depth $L_{THz}$ which depends on the carrier density in the sample. The reflected THz beam is collected by a silicon beam-splitter and focused on a 1 mm-thick (110) GaP crystal. The THz beam propagates in a box purged with dry air to avoid absorption by water molecules.

The gate beam is focused on the same GaP crystal to perform electro-optic sampling detection of the THz electric field, using a λ/4 waveplate, a Wollaston prism and two balanced silicon photodiodes connected to a lock-in and a DAQ card. An optical chopper is set on the THz path at 500 Hz (synchronized to the 1kHz-laser) for the lock-in detection.

Hereafter, the sampling time between the THz and the gate pulse is called "time t". It is varied using a delay line in the gate beam path. The delay between the optical pump and the THz probe is called "the pump-probe delay $\tau$". It is varied using a delay line in the pump beam path. We set $\tau = 0$, for the delay which gives the maximum reflected THz amplitude.

The InSb sample is a 500µm-thick bulk wafer, (100) oriented and commercially available. The sample was nominally undoped but a residual extrinsic doping concentration of ~$10^{14}$ cm$^{-3}$ is estimated from previous measurements of the plasma frequency at $f_p \sim 0.2$ THz[23]. The sample was cooled down to 77 K in a cryostat equipped with TPX (polymer) windows.

3. Results

The temporal traces of the THz pulse reflected from InSb, measured at the pump fluence of 31 µJ/cm² for different pump-probe delays τ (color lines), are plotted in figure 2.a, with zooms on the extrema as insets. We observe an increase of the reflected THz peak-to-peak amplitude as the optical pump excites the sample, which then decreases back towards equilibrium as the pump-probe delays τ increases. We notice that the peaks shift in time t for varying delay τ, as the shape of the THz field is significantly modified.

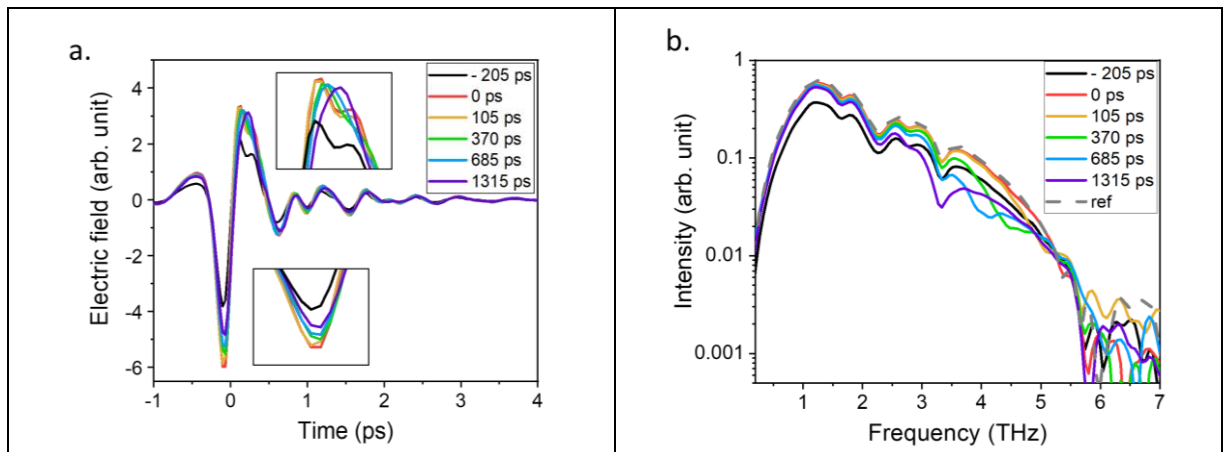

Figure 2: a. Temporal traces of the THz pulse reflected from InSb sample, at different pump-probe delays τ (color lines) after the pump excitation with the fluence of 31 µJ/cm². Inset: zoom on both positive and negative peaks. b. THz spectra of the THz temporal traces shown in the panel a. The reference spectrum, acquired on InSb sample pumped at a higher fluence of 300 µJ/cm² at τ ~ 0 ps, is plotted in grey dashed line.

For the THz temporal traces, we applied a Gaussian filtering to reduce the Fabry-Perot effect due to internal reflections in the BNA crystal/substrate, and performed Fourier transforms to obtain the reflected spectra presented in figure 2.b. The whole spectrum intensity increases as the pump excites the sample (τ = 0 ps), then progressively decreases for high frequencies first, followed by lower frequencies as τ > 500 ps.

We derive the reflectivity from the ratio between the pumped spectrum at a given τ and the reference spectrum. The reference spectrum was measured on the pumped sample at a very short delay after the pump arrival (τ ~ 0 ps), for a much higher fluence of 300 µJ/cm². At this fluence, assuming that one absorbed photon excites one electron-hole pair, and considering the optical penetration depth of 380 nm at the pump wavelength of 1030 nm and the reflection coefficient of R = 0.38, we estimated the density of the photoexcited carriers at $2 \cdot 10^{19}$ cm$^{-3}$ (giving a plasma frequency of 85 THz >> 5 THz). Using this "photo-screening method"[24], a transient metallic state is briefly induced that intrinsically sits at the same position of the sample, so should act as a relevant reference in reflection. The reference spectrum, plotted in dashed gray line in figure 2b, was measured just before the pump-probe scans, to ensure similar conditions.

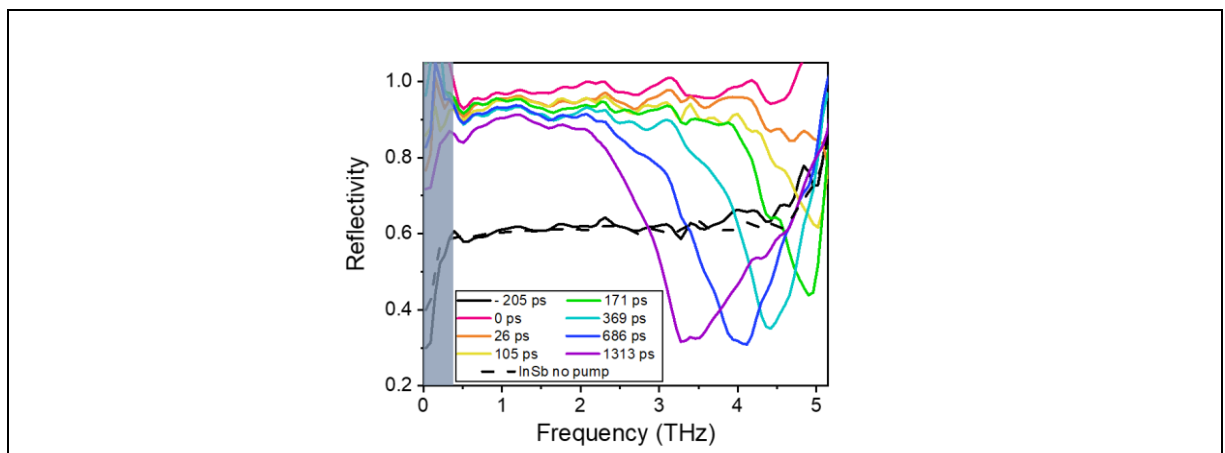

Figure 3: Reflectivity spectra for different pump-probe delays τ at a fluence of 31 µJ/cm². The plain black line depicts the reflectivity spectrum measured at the negative pump-probe delay -205 ps, while the dashed black line corresponds to the reflectivity spectrum measured in the absence of the optical excitation. The gray area is not reliable, resulting from a low signal-to-noise in this spectral range.

The reflectivity spectra obtained for different pump-probe delays τ at the pump fluence of 31 µJ/cm² are presented in figure 3. A good match between the reflectivity signal measured at negative pump-probe delays and in the absence of the pump pulse indicates that the sample recovers completely after the optical excitation and there is no cumulative effect. The equilibrium reflectivity behaves as expected for nominally undoped InSb sample. The plasma frequency is expected at 0.2 THz, but it is not visible in these measurements because of the noise level. Then there is a plateau with the reflectivity value of 0.6, and an increase reaching the reflectivity value of 1 close to 5 THz due to the optical phonon contribution ($f_{LO} = 5.71$ THz, $f_{TO} = 5.34$ THz[25]).

Just after the optical excitation at short delays τ, the reflectivity rises up to almost 1 on the full spectral range as a large carrier density is photo-generated (red, orange lines). The plasma frequency, defined later in equation (2), significantly increases up to frequencies way above the detected spectral range. For the pump fluence of 31 µJ/cm² and the penetration depth of 380 nm (for λ = 1030 nm), we estimate that the pump induces a carrier density about 2,5.10$^{18}$ cm$^{-3}$ corresponding to the plasma frequency of 30 THz. For longer delays τ, the carrier diffusion and electron-hole pair recombination lead to a decrease of the carrier density. An edge appears in the measured spectral range, corresponding to the plasma frequency, which shifts progressively towards lower frequencies, down to ~3 THz at τ ~ 1 ns.

The sample is not yet back to equilibrium at the maximum τ delay accessible with the delay line (purple line), as InSb is well-known for its very long recombination time[26]. From an Auger coefficient[27] of $7.10^{-26}$ cm$^6$.s$^{-1}$, we estimate the recombination time in the range of [1-100] ns for carrier densities of [10$^{16}$ - 10$^{17}$] cm$^{-3}$, which is much smaller than 1 ms delay between two pulses at 1kHz repetition rate. We notice that the reflectivity below 2 THz seems to decrease from 0.96 to 0.9, which could already be an indication of the THz pulse probing an inhomogeneously pumped volume as discussed in more details below. However, the photo-screening method with one reference acquired just before the measurement may also have limitations. It is possible that we are measuring in a regime where screening is not fully achieved for high frequencies, especially around the phonon frequency at 5 THz.

4. Weighted Drude-Lorentz model of the reflectivity

The THz response resulting from an optical pump pulse absorbed by InSb sample has been previously described[8]. Using the Drude-Lorentz model, the complex dielectric constant $\mathcal{E}(f)$ of excited InSb can be expressed as:

$$\widehat{\mathcal{E}}(f,N) = \mathcal{E}_\infty \left(1 - \frac{f_p^2(N)}{f^2 + i\gamma(N)f} + \frac{f_{LO}^2 - f_{TO}^2}{f_{TO}^2 - f^2}\right) \quad (1)$$

where $\mathcal{E}_\infty$ is the high frequency permittivity equal to 15.68 in InSb [1], $\gamma(N)$ is the carrier-dependent scattering rate, computed from the carrier density dependent mobility data at 77 K from reference [28] and $f_p$ is the electron plasma frequency:

$$f_p^2(N) = \frac{1}{(2\pi)^2} \frac{e^2 N}{\mathcal{E}_0 \mathcal{E}_\infty m_e^*(N)} \quad (2)$$

e is the elementary charge and $m_e^*(N)$ the carrier-density-dependent effective mass of electrons in the conduction band, due to the non-parabolicity of InSb[29]. Only electrons are considered in eq. (1). Despite their similar band masses, light holes (LH) do not significantly contribute to the conduction. Indeed, the LH band is hardly populated due to the presence of a more favorable heavy hole band. The heavy holes do not contribute either because of their larger effective mass.

From the dielectric function expression, one can calculate the complex refractive index such as $\hat{n}(f)^2 = \hat{\varepsilon}(f)$ and the reflectivity at the normal incidence:

$$R(f) = \left|\frac{1-\hat{n}(f)}{1+\hat{n}(f)}\right|$$

In equilibrium, the sample has a reflectivity $R_{eq}(f, N_0)$, where $N_0$ is the extrinsic doping level $\sim 10^{14}$ cm$^{-3}$ corresponding to the equilibrium plasma frequency $f_p(N_0) \sim 0.2$ THz.

Besides the Drude-Lorentz description, the mismatch between the pump and probe penetration should be considered. After the absorption of the pump pulse, optically excited carriers will diffuse and recombine in the material. We assume that for a given delay τ, there is a homogeneous electron density $N(\tau)$, over a distance $L_{diff}(\tau)$, which results from both diffusion and recombination processes. This excited layer appears in the total THz reflectivity with a contribution $R_{exc}(f, N(\tau))$.

The total THz reflectivity results from the volume probed by the THz pulse, whose penetration depth will depend on the carrier density. The penetration depth is defined at a given delay τ :

$$L_{THz}(f, N(\tau)) = \frac{c}{2\pi f . Im(\sqrt{\hat{\varepsilon}(f, N(\tau))}}$$

with c the velocity of light.

Since our bandwidth is ~ 5 THz and its spectral weight is mostly centered on [1-2] THz, we fix the frequency at 1.5 THz and approximate $L_{THz}(N(\tau)) = L_{THz}(f = 1.5\ THz, N(\tau))$.

Finally, in this one-layer model which approximates the profile of the photo-excited carriers and the THz electric field with step-functions, we define the total THz reflectivity as an average between the excited reflectivity and equilibrium reflectivity, with weights corresponding to the relative diffusion and penetration lengths:

$$R_{tot}(f,\tau) = \frac{L_{diff}(\tau)}{L_{THz}(N(\tau))} R_{exc}(f, N(\tau)) + \frac{L_{THz}(N(\tau)) - L_{diff}(\tau)}{L_{THz}(N(\tau))} R_{eq}(f, N_0) \qquad (3)$$

with the condition $L_{diff}(\tau) \leq L_{THz}(N(\tau))$ in the model.

Figure 4 presents the experimental reflectivity spectra for 3 fluence values, at different pump-probe delays and the corresponding fitting curves, using the model in equation (3). $N(\tau)$ and $L_{diff}(\tau)$ are free parameters, $\gamma(N)$ and $m^*(N)$ were constrained to experimental values from the mobility data in ref [28] and effective mass data from ref [30], after estimating N values from the first fit iteration.

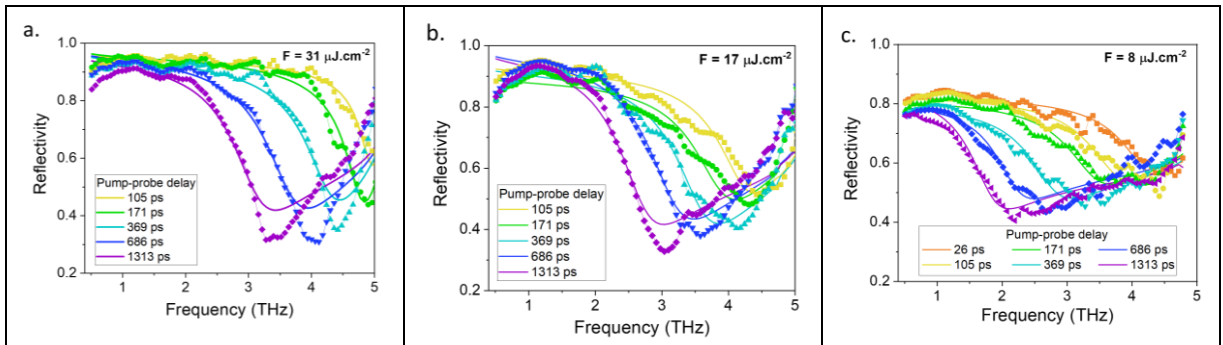

Figure 4: a. b. c. Reflectivity spectra (dots + thin line) and corresponding fitting curves (thick plain line) using equation 3, for different pump-probe delays (color lines), respectively for 31 µJ/cm², 17 µJ/cm² and 8 µJ/cm².

For very short pump-probe delays $\tau$, the plasma edge is shifted way above the measured spectral range. Given our signal-to-noise ratio, the reflectivity at these short delays cannot be reliably approached with the Drude-Lorentz model, as the model leads to high uncertainty on the extracted plasma frequency values. The fitting curves at larger delays are in overall good agreement with the experimental data, with R-square values always > 0.91. At the fluence of 8 µJ/cm², the plasma edge does not exceed the limit of the accessible spectral range, even for short pump-probe delays, and the reflectivity below 2 THz remains below 0.85, as a hint for penetration mismatch. Indeed, the lower the pump fluence, the smaller the excited carrier density and the larger the THz penetration depth. It leads to the situation where the THz pulse probes both excited and unperturbed parts of the sample.

5. Discussion

From the fit parameters, we extract $N(\tau)$ which is a crucial parameter to characterize the carrier dynamics. The carrier density values as a function of $\tau$ are plotted in figure 5 for 3 fluence values (squares). For a given pump-probe delay, the photo-excited carrier density increases with increasing fluence, as more carrier are photo-generated. In the fluence regime we explore, we do not notice any saturation effect as it was observed in other works[8,14]. We propose to model $N(\tau)$ with exponential decay fits of the form ($Ae^{-\frac{\tau}{T}} + B$) to estimate recombination time constants at different fluences. We obtain the following values $T = 131 \pm 75$ ps, $T = 193 \pm 115$ ps and $T = 150 \pm 30$ ps, respectively for 31 µJ/cm², 17 µJ/cm² and 8 µJ/cm² pump fluences. With the limited number of pump-probe delays in this study, the uncertainty is quite high, so we only conclude that the time constants are on the same order in the examined fluence range. These values are reasonably consistent, although somewhat shorter with respect to the time constant values from previous OPTP work at fixed gate time[14], where they obtained a time constant of 560 ps for the pump fluence of 32 µJ/cm², pumping at 800 nm. A more quantitative comparison would require more data points to refine our estimation of the recombination time from $N(\tau)$.

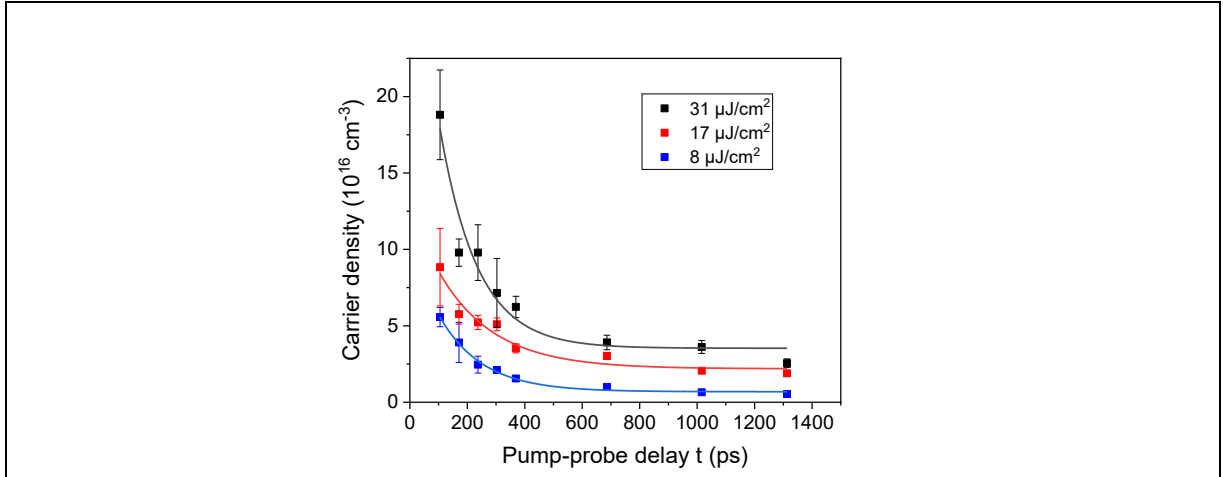

Figure 5: Experimental carrier density as a function of pump-probe delays for 31 µJ/cm² (black squares), 17 µJ/cm² (red squares) and 8 µJ/cm² (blue squares). Exponential fits (plain lines) of the form $Ae^{-\frac{\tau}{T}} + B$ are proposed to approach the carrier density values extracted from the data.

From the fit parameters, we also extract $L_{diff}(N(\tau))$ which compares with the THz penetration depth $L_{THz}(N(\tau))$ to set the ratio of the excited part over the equilibrium part of the sample that is actually probed by the THz pulse. For further comparison, we also calculate a theoretical value for the diffusion length $L_{diff,th}(N(\tau))$, for given $\tau$ and $N(\tau)$, since we can estimate the diffusion coefficient from the N-dependent mobility $\mu(N)$[28] and the Einstein relation[31]:

$$D(\tau) = \frac{\mu(N(\tau))k_B T}{e} = \frac{L_{diff,th}(N(\tau))^2}{\tau}$$

where $k_B$ is the Boltzmann constant, D is the diffusion coefficient, and T is the temperature (77 K).

The comparison between the THz penetration depth $L_{THz}(N(\tau))$, the fit parameter $L_{diff}(N(\tau))$ and the theoretical value $L_{diff,th}(N(\tau))$, is plotted in figure 6, as a function of $N(\tau)$. As expected, $L_{THz}(N(\tau))$ always increases with decreasing $N(\tau)$, the THz field penetrates more in the sample as the carrier density decreases.

For the two highest fluences, we obtain $L_{diff}(N(\tau)) \sim L_{THz}(N(\tau))$, meaning that the volume probed by the THz pulse is completely excited, for every pump-probe delay. In that respect, $L_{diff}(N(\tau))$ is an estimated minimum value for the carrier diffusion length, as the THz potentially probes only a portion of the excited part of the sample where carriers have diffused. This is confirmed with the comparison to the theoretical value for the diffusion length. If at high carrier density, $L_{diff}(N(\tau))$ is close to $L_{diff,th}(N(\tau))$ as both THz penetration length and diffusion length are small, at lower carrier density (longer pump-probe delays), the fit values are much smaller than the theoretical values, limited by the THz penetration depth to estimate a minimum for the diffusion length.

For the fluence of 8 µJ/cm², we obtain from the fit that the ratio of the excited part over the equilibrium part of the sample that is actually probed by the THz pulse is between 58% and 68% depending on delays, as the THz pulse penetrates more into the material at lower carrier density. At this fluence, $L_{diff}(N(\tau))$ fit estimations are close to the theoretical diffusion length values, as shown in figure 6, since the THz pulse probes the total excited volume and some volume in equilibrium. In this case, our model can provide a reliable estimation of the carrier diffusion length.

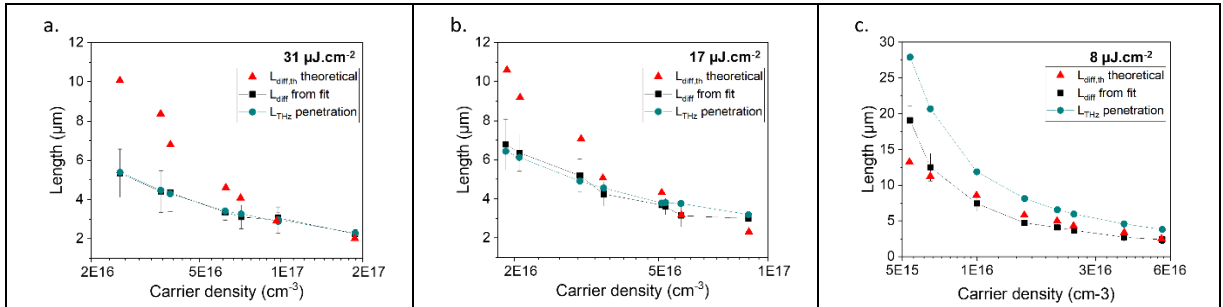

Figure 6: Comparison of THz penetration length $L_{THz}(N(\tau))$, (green dots), estimated diffusion length from the model $L_{diff}(N(\tau))$ (black squares) and theoretical value for the diffusion length $L_{diff,th}(N(\tau))$ (red triangles), as a function of the estimated carrier density $N(\tau)$, respectively for a. 31 µJ/cm², b. 17 µJ/cm² and c. 8 µJ/cm².

6. Conclusion

In summary, we have implemented an OPTP setup, based on broadband THz generation from optical rectification in an organic BNA crystal. The setup was used to investigate the photo-induced carrier dynamics in a semiconductor sample of InSb. Using the Drude-Lorentz model with weights to account for length mismatches between the THz penetration length and the excited sample length, we extracted an estimate of the carrier density and the carrier diffusion length for each pump-probe delay $\tau$, in reasonable agreement with other works and with theoretical values, respectively.

We showed that the spectral method enables the absolute retrieval of carrier density dynamics[8] and avoids artifacts caused by THz temporal trace shifts when performing measurements at a fixed gate delay only. To refine the results, we could increase our signal-to-noise ratio to reliably model the response at very short pump-probe delays, even if the plasma edge is pushed over the limit of the spectral range[32]. The photo-screening method for referencing has been conclusive in this work, supported by previous studies[24], but it could still be improved by measuring more systematically the reference spectra with automated flipping optics. Our simple step model provided insights on the mismatch between the THz penetration depth and the actual excited sample depth after carrier diffusion, whose evolutions with time and pump fluence are not intuitive as both quantities depend on carrier density. With the facilitated access to broadband THz sources including spintronic emitters and organic crystals, this work reinforces that the systematic exploitation of the spectral information provides a complete picture of the carrier dynamics.

Acknowledgements

The authors thank Yannis Laplace for insightful discussions and Martial Nicolas for his support on the mechanical development of the setup. We acknowledge funding from the Agence National de la Recherche via the grants ANR "SUPER2DTMD" and ANR "TERADIRAC". This work has been supported by Region Ile-de-France in the framework of DIM QuanTiP and DIM SIRTEQ.

The data that support the findings of this study are available from the corresponding author upon reasonable request.

References

[1] F. D'Angelo, Z. Mics, M. Bonn, and D. Turchinovich, "Ultra-broadband THz time-domain spectroscopy of common polymers using THz air photonics," Opt. Express **22**(10), 12475–12485 (2014).

[2] A. Bera, S. Bera, S. Kalimuddin, S. Gayen, M. Kundu, B. Das, and M. Mondal, "Review of recent progress on THz spectroscopy of quantum materials: superconductors, magnetic and topological materials," The European Physical Journal Special Topics **230**(23), 4113–4139 (2021).

[3] R. Ulbricht, E. Hendry, J. Shan, T.F. Heinz, and M. Bonn, "Carrier dynamics in semiconductors studied with time-resolved terahertz spectroscopy," Rev. Mod. Phys. **83**(2), 543–586 (2011).

[4] S. Mitra, L. Avazpour, and I. Knezevic, "Terahertz conductivity of two-dimensional materials: a review," Journal of Physics: Condensed Matter **37**(13), 133005 (2025).

[5] J.A. Spies, J. Neu, U.T. Tayvah, M.D. Capobianco, B. Pattengale, S. Ostresh, and C.A. Schmuttenmaer, "Terahertz Spectroscopy of Emerging Materials," J. Phys. Chem. C **124**(41), 22335–22346 (2020).

[6] H.-K. Nienhuys, and V. Sundström, "Intrinsic complications in the analysis of optical-pump, terahertz probe experiments," Phys. Rev. B **71**(23), 235110 (2005).

[7] Tim Suter, Zia Macdermid, Zekai Chen, Steven Lee Johnson, and Elsa Abreu, "Terahertz time-domain spectroscopy of materials under high pressure in a diamond anvil cell," (n.d.).

[8] P. Fischer, G. Fitzky, D. Bossini, A. Leitenstorfer, and R. Tenne, "Quantitative analysis of free-electron dynamics in InSb by terahertz shockwave spectroscopy," Phys. Rev. B **106**(20), 205201 (2022).

[9] Samira Mansourzadeh, Tim Vogel, Alan Omar, Megan F. Biggs, Enoch S.-H. Ho, Claudius Hoberg, David J. Michaelis, Martina Havenith, Jeremy A. Johnson, and Clara J. Saraceno, "High-Dynamic Range Broadband Terahertz Time-Domain Spectrometer Based on Organic Crystal MNA," (n.d.).

[10] Z.B. Zaccardi, I.C. Tangen, G.A. Valdivia-Berroeta, C.B. Bahr, K.C. Kenney, C. Rader, M.J. Lutz, B.P. Hunter, D.J. Michaelis, and J.A. Johnson, "Enabling high-power, broadband THz generation with 800-nm pump wavelength," Opt. Express **29**(23), 38084–38094 (2021).

[11] S. Mansourzadeh, T. Vogel, M. Shalaby, F. Wulf, and C.J. Saraceno, "Milliwatt average power, MHz-repetition rate, broadband THz generation in organic crystal BNA with diamond substrate," Opt. Express **29**(24), 38946–38957 (2021).

[12] S.C. Howells, and L.A. Schlie, "Transient terahertz reflection spectroscopy of undoped InSb from 0.1 to 1.1 THz," Applied Physics Letters **69**(4), 550–552 (1996).

[13] J. Chochol, K. Postava, M. Čada, M. Vanwolleghem, L. Halagačka, J.-F. Lampin, and J. Pištora, "Magneto-optical properties of InSb for terahertz applications," AIP Advances **6**(11), 115021 (2016).

[14] G. Li, W. Zhou, W. Zhang, G. Ma, H. Cui, Y. Gao, Z. Huang, and J. Chu, "Pump fluence dependence of ultrafast carrier dynamics in InSb measured by optical pump–terahertz probe spectroscopy," Appl. Opt. **57**(33), 9729–9734 (2018).

[15] C.Q. Xia, M. Monti, J.L. Boland, L.M. Herz, J. Lloyd-Hughes, M.R. Filip, and M.B. Johnston, "Hot electron cooling in InSb probed by ultrafast time-resolved terahertz cyclotron resonance," Phys. Rev. B **103**(24), 245205 (2021).

[16] M.C. Hoffmann, J. Hebling, H.Y. Hwang, K.-L. Yeh, and K.A. Nelson, "Impact ionization in InSb probed by terahertz pump---terahertz probe spectroscopy," Phys. Rev. B **79**(16), 161201 (2009).

[17] S. Houver, L. Huber, M. Savoini, E. Abreu, and S.L. Johnson, "2D THz spectroscopic investigation of ballistic conduction-band electron dynamics in InSb," Opt. Express **27**(8), 10854–10865 (2019).

[18] H. Tanimura, J. Kanasaki, and K. Tanimura, "Ultrafast scattering processes of hot electrons in InSb studied by time- and angle-resolved photoemission spectroscopy," Phys. Rev. B **91**(4), 045201 (2015).

[19] R.K. Lal, and P. Chakrabarti, "A comparison of dominant recombination mechanisms in n-type InAsSb materials," Progress in Crystal Growth and Characterization of Materials **52**(1), 33–39 (2006).

[20] R.D. Grober, H.D. Drew, G.L. Burdge, and B.S. Bennett, "Direct measurement of the recombination rates in bulk InSb by time-resolved photoluminescence," Journal of Applied Physics **71**(10), 5140–5145 (1992).

[21] G. Li, X. Nie, W. Zhou, W. Zhang, H. Cui, N. Xia, Z. Huang, J. Chu, and G. Ma, "Influence of doping for InSb on ultrafast carrier dynamics measured by time-resolved terahertz spectroscopy," Appl. Opt. **59**(35), 11046–11052 (2020).

[22] "Terahertz Innovations," (n.d.).

[23] I. Aupiais, R. Grasset, T. Guo, D. Daineka, J. Briatico, S. Houver, L. Perfetti, J.-P. Hugonin, J.-J. Greffet, and Y. Laplace, "Ultrasmall and tunable TeraHertz surface plasmon cavities at the ultimate plasmonic limit," Nature Communications **14**(1), 7645 (2023).

[24] Lucas Killian Karl Huber, Nonlinear Electromagnetic Probes for the Study of Ultrafast Processes in Condensed Matter, ETH Zürich, 2017.

[25] Peter Y. Yu, Manuel Cardonna, *Fundamentals of Semicondcutors: Physics and Materials Properties*, Springer Berlin, Heidelberg (2010).

[26] R.A. Laff, and H.Y. Fan, "Carrier Lifetime in Indium Antimonide," Phys. Rev. **121**(1), 53–62 (1961).

[27] S Marchetti, M Martinelli, and R Simili, "The InSb Auger recombination coefficient derived from the IR-FIR dynamical plasma reflectivity," Journal of Physics: Condensed Matter **13**(33), 7363 (2001).

[28] E. Litwin-Staszewska, W. Szymańska, and R. Piotrzkowski, "The Electron Mobility and Thermoelectric Power in InSb at Atmospheric and Hydrostatic Pressures," Physica Status Solidi (b) **106**(2), 551–559 (1981).

[29] S. Zollner, C.A. Armenta, S. Yadav, and J. Menéndez, "Conduction band nonparabolicity, chemical potential, and carrier concentration of intrinsic InSb as a function of temperature," Journal of Vacuum Science & Technology A **43**(1), 012801 (2024).

[30] I.P. Molodyan, D.N. Nasledov, S.I. Radautsan, and V.G. Sidorov, "The effective mass of electrons in (InSb)x·(InTe)1−x crystals," Physica Status Solidi (b) **18**(2), 677–682 (1966).

[31] A.H. Marshak, and D. Assaf, "A generalized Einstein relation for semiconductors," Solid-State Electronics **16**(6), 675–679 (1973).

[32] J. Guise, H. Ratovo, M. Thual, P. Fehlen, F. Gonzalez-Posada Flores, J.-B. Rodriguez, L. Cerutti, E. Centeno, S. Blin, and T. Taliercio, "Measuring low doping level and short carrier lifetime in indium arsenide with a contactless terahertz technique at room temperature," Journal of Applied Physics **134**(16), 165701 (2023).